\documentclass[aps,pre,final,10pt,twocolumn,superscriptaddress,nofootinbib,flushbottom,showpacs]{revtex4-1}% 
\usepackage{amssymb,amsmath,amsfonts}
\usepackage{graphicx,color}
\usepackage[squaren]{SIunits}
\usepackage[]{units}
\usepackage[english]{babel}
\usepackage{tabularx}

\newcommand{\dif}{\mathrm{d}}%
\newcommand{\Sp}{{\mathcal{S}}}%
\newcommand{\Sphere}{{\mathbb{S}_{2}}}%

\newcommand{\rb}{\mathbf{r}}%
\newcommand{\pb}{\mathbf{p}}%
\newcommand{\pbc}{\hat{p}}%
\newcommand{\dpbc}{\hat{\delta p}}%
\newcommand{\pp}{\mathfrak{p}}%
\newcommand{\PP}{\mathcal{P}}%
\newcommand{\qb}{\mathbf{q}}%
\newcommand{\qbc}{\hat{q}}%
\newcommand{\uu}{\mathbf{\hat{u}}}%
\newcommand{\vb}{\mathbf{v}}%
\newcommand{\vm}{{v_{m}}}%
\newcommand{\vthmin}{v_{\mathrm{th,min}}}%
\newcommand{\alphathmin}{\alpha_{\mathrm{th,min}}}%

\newcommand{\norm}[1]{\lVert#1\rVert}%
\newcommand{\R}{\mathbb{R}}%
\newcommand{\C}{\mathbb{C}}%

\newcommand{\Nsh}{n}%
\newcommand{\Nth}{N_{\theta}}%
\newcommand{\Nphi}{N_{\phi}}%
\newcommand{\TempR}{{\varepsilon}}%

\newcommand{\cn}{\zeta}%
\newcommand{\nn}{\mathfrak{N}}%

\newcommand{\Tang}[2]{{\mathrm{T}_{#1}{#2}}}%
\newcommand{\TS}{{\Tang{\rb}{\Sp}}}%

\makeatletter
\newcommand*\bigcdot{\mathpalette\bigcdot@{.5}}
\newcommand*\bigcdot@[2]{\mathbin{\vcenter{\hbox{\scalebox{#2}{$\m@th#1\bullet$}}}}}
\makeatother

\newcommand{\Grad}{{\operatorname{grad}}}%
\newcommand{\Div}{{\operatorname{div}}}%
\newcommand{\GradS}{{\operatorname{grad}_{\Sp}}}%
\newcommand{\rotS}{{\operatorname{rot}_{\Sp}}}%
\newcommand{\RotS}{{\operatorname{Rot}_{\Sp}}}%
\newcommand{\DivS}{{\operatorname{div}_{\Sp}}}%
\newcommand{\Laplace}{\boldsymbol{\triangle}}%
\newcommand{\LaplaceS}{\Laplace_{\Sp}}%
\newcommand{\LaplaceDR}{\Laplace_{\mathrm{dR}}}%

\newcommand{\Tstrut}{\rule{0pt}{2.6ex}}%
\newcolumntype{L}{>{$}l<{$}}%
\newcolumntype{R}{>{$}r<{$}}%
\newcolumntype{Y}{>{\centering\arraybackslash}X}%

\begin{document}

\author{Simon \surname{Praetorius}}
\email[Corresponding author: ]{simon.praetorius@tu-dresden.de}
\affiliation{Institute for Scientific Computing, Technische Universit\"{a}t Dresden, D-01062 Dresden, Germany}

\author{Axel \surname{Voigt}}
\affiliation{Institute for Scientific Computing, Technische Universit\"{a}t Dresden, D-01062 Dresden, Germany}
\affiliation{Dresden Center for Computational Materials Science (DCMS), D-01062 Dresden, Germany}
\affiliation{Center for Systems Biology Dresden (CSBD), D-01307 Dresden, Germany}

\author{Raphael \surname{Wittkowski}}
\affiliation{Institut f\"{u}r Theoretische Physik, Westf\"alische Wilhelms-Universit\"{a}t M\"{u}nster, D-48149 M\"{u}nster, Germany}
\affiliation{Center for Nonlinear Science (CeNoS), Westf\"alische Wilhelms-Universit\"{a}t M\"{u}nster, D-48149 M\"{u}nster, Germany}

\author{Hartmut \surname{L\"owen}}
\affiliation{Institut f\"{u}r Theoretische Physik II: Weiche Materie, Heinrich-Heine-Universit\"{a}t D\"{u}sseldorf, D-40225 D\"{u}sseldorf, Germany}

\title{Active crystals on a sphere}
\date{\today}

\begin{abstract}
Two-dimensional crystals on curved manifolds exhibit nontrivial defect structures. Here, we consider ``active crystals'' on a sphere, which are composed of self-propelled colloidal particles. Our work is based on a new phase-field-crystal-type model that involves a density and a polarization field on the sphere. Depending on the strength of the self-propulsion, three different types of crystals are found: a static crystal, a self-spinning ``vortex-vortex'' crystal containing two vortical poles of the local velocity, and a self-translating ``source-sink'' crystal with a source pole where crystallization occurs and a sink pole where the active crystal melts. These different crystalline states as well as their defects are studied theoretically here and can in principle be confirmed in experiments.
\end{abstract}


% Colloids: 82.70.Dd
% Point defects and defect clusters: 61.72.J-
% Finite-element and Galerkin methods: 02.70.Dh
% 05.70.Fh 	Phase transitions
% 05.70.Ln 	Nonequilibrium and irreversible thermodynamics
% 61.72.-y 	Defects and impurities in crystals; microstructure
% 61.72.Bb 	Theories and models of crystal defects
% 61.72.Cc 	Kinetics of defect formation and annealing

\maketitle

\section{\label{sec:introduction}Introduction}
It is common wisdom that the plane can be packed periodically by hexagonal crystals of spherical particles but, when the manifold is getting curved, defects emerge due to topological constraints. The most common example is a soccer ball that has a tiling of hexagons and pentagons. Indeed, similar structures are realized by Wigner-Seitz cells in particle layers covering a sphere, which is a topic that has been recently explored a lot in physics (for reviews see Refs.\ \cite{Nelson2002,Bowick2009}). Mathematically this topic is related to the classical problem of finding the minimal energy distribution of interacting points on a sphere \cite{Smale98,Backofen2011}. Likewise, while a unit vector field can be uniform in flat space, it is well-known that ``a hedgehog cannot be combed in a continuous way'' \cite{AgricolaF2002}, which results in topological defects of an oriented vector field on a sphere.

Recently, also self-propelled (i.e., ``active'') colloidal particles, which dissipate energy while they move, have been studied a lot \cite{Ramaswamy2010,Romanczuk2012,Elgeti2015,Menzel2015,Bechinger2016}. At large density in the plane, these particles form crystals under nonequilibrium conditions  \cite{Bialke2012,Redner2013,MenzelL2013,Ferrante2013,Menzel2014,Briand2016}. Self-propelled particles can also be confined to a compact manifold like a sphere, as realized by multicellular spherical Volvox colonies \cite{Drescher2010}, bacteria moving on oil drops \cite{Juarez2014} or layered in water drops \cite{Vladescu2014}, or by active nematic vesicles \cite{Keber2014}. This has triggered recent theoretical and simulation work on self-propelled particles on spheres considering both their individual \cite{Grossmann2015,Li2015} and collective \cite{Sknepnek2015,Alaimo2017} dynamics.

Here, we unify the two fields of equilibrium crystals and self-propelled colloidal particles on curved manifolds and study an active crystal on a sphere. For this purpose, we use a phase-field-crystal-type model \cite{Elder2002,Elder2004,Teeffelen2009,EmmerichEtAl2012}, which we obtain by generalizing a previously proposed phase-field-crystal (PFC) model for active crystals in the plane \cite{MenzelL2013,Menzel2014} to the sphere. The model involves both a scalar density field and a polarization vector field on the sphere. Depending on the strength of the self-propulsion, three different crystalline states are found:
i) a static crystal similar to its equilibrium counterpart, 
ii) a self-spinning ``vortex-vortex'' crystal, which contains two vortical poles of the local velocity field, and
iii) a self-translating ``source-sink'' crystal, which has a pole of the local velocity field where crystallization occurs (``source'') as well as one where the active crystal melts (``sink''). 
Our work goes beyond recent studies on active nematic shells where the density field is homogeneous \cite{KhoromskaiaA2017}, Toner-Tu-like models on curved spaces that cannot describe crystalline states \cite{Fily2016,Shankar2017}, and a combination of an equilibrium crystal on a sphere with a single self-propelled tracer particle \cite{Yao2016}.

This article is organized as follows: We describe our PFC model for active crystals on a sphere in Sec.\ \ref{sec:surface_pfc} and the numerical solution of the associated equations in Sec.\ \ref{sec:numerical_solution}. The results that we obtained by numerically solving this PFC model are presented in Sec.\ \ref{sec:results}. Finally, we conclude in Sec.\ \ref{sec:conclusions}.

\section{\label{sec:surface_pfc}A phase-field-crystal model for active crystals on a sphere}
In the plane, active colloidal crystals can be described by a rescaled density field $\psi(\rb,t)$, which we simply call ``density field'' in the following, and a polarization field $\pb(\rb,t)$, where $\rb$ and $t$ denote position and time, respectively. While $\psi(\rb,t)$ describes the spatial variation of the particle number density at time $t$, $\pb(\rb,t)$ describes the time-dependent local polar order of the particles. 
Using suitably scaled units of length, time, and energy, a minimal field-theoretical model for active crystals in the plane is given by \cite{MenzelL2013,Menzel2014}
{\allowbreak\begin{align}%
\partial_{t} \psi &= \Laplace\frac{\delta\mathcal{F}}{\delta\psi} - v_{0}\,\Div\pb \,, \label{eq:apfc_planeI} \\
\partial_{t} \pb &= (\Laplace - D_{r})\frac{\delta\mathcal{F}}{\delta\pb} - v_{0}\,\Grad\psi \,. \label{eq:apfc_planeII}
\end{align}}%
This PFC model can describe crystallization in active systems on microscopic length and diffusive time scales. 
Here, $\partial_{t}=\partial/\partial t$ denotes a partial time derivative, $\Laplace$ is the ordinary Cartesian Laplace operator, $\delta/\delta\psi$ and $\delta/\delta\pb$ are functional derivatives with respect to $\psi$ and $\pb$, respectively, and $\Grad$ and $\Div$ are the ordinary Cartesian gradient and divergence operators, respectively.
$v_{0}$ is an activity parameter that describes the self-propulsion speed of the active colloidal particles \cite{MenzelL2013,Alaimo2016} and $D_r$ is their rescaled rotational diffusion coefficient.
Furthermore, $\mathcal{F}[\psi,\pb]=\mathcal{F}_{\psi}[\psi]+\mathcal{F}_{\pb}[\pb]$ is a free-energy functional with the traditional PFC functional \cite{Elder2002,Elder2004} 
\begin{equation}
\mathcal{F}_{\psi}[\psi] = \int_{\R^{2}}\!\! \Big( \frac{1}{2}\psi\big(\TempR + (1 + \Laplace)^{2}\big)\psi + \frac{1}{4}\psi^{4} \Big) \dif^{2}r 
\label{eq:F_psi}%
\end{equation}
and the polarization-dependent contribution \cite{MenzelL2013,Menzel2014}
\begin{equation}
\mathcal{F}_{\pb}[\pb] = \int_{\R^{2}}\!\! \Big( \frac{1}{2} C_{1} \norm{\pb}^{2} + \frac{1}{4}C_{2} \norm{\pb}^{4} \Big) \dif^{2}r \,,  
\label{eq:F_p}%
\end{equation}
where $\norm{\,\cdot\,}$ is the Euclidean norm. 
The constant $\TempR$ sets the temperature \cite{Elder2002,Elder2004} and the coefficients $C_{1}$ and $C_{2}$ affect the local orientational ordering due to the drive of the particles. While $C_{1}$ takes diffusion of the polarization field into account and should be positive, $C_{2}$ describes a higher-order contribution that can be neglected when studying active crystals \cite{MenzelL2013}. 
In contrast to the traditional PFC model \cite{Elder2002}, there is not the wave number $k_{0}$ preferred by the system as an additional parameter in Eq.\ \eqref{eq:F_psi}. We set $k_{0} = 1$, thus the preferred lattice constant is $2\pi$ in the chosen dimensionless units.

To describe active crystals on a sphere $\Sp=R\,\Sphere$ with radius $R$, where $\Sphere$ is the three-dimensional unit sphere, we start from the PFC model for the plane given by Eqs.\ \eqref{eq:apfc_planeI}-\eqref{eq:F_p} and extend it appropriately. First, we parametrize the position $\rb$, which becomes a three-dimensional vector that describes positions on the sphere $\Sp$, by $\rb(\theta,\phi)=R \uu(\theta,\phi)$ with the orientational unit vector $\uu(\theta,\phi)=(\sin(\theta)\cos(\phi),\sin(\theta)\sin(\phi),\cos(\theta))^{\mathrm{T}}$ and the spherical coordinates $\theta\in [0,\pi]$ and $\phi\in [0,2\pi)$.
Next, we define the polarization field $\pb(\rb,t)$ as a three-dimensional vector field that is tangential to $\Sp$ at $\rb$, i.e., $\pb(\rb,t)=p_{\theta}(\rb,t)\partial_{\theta}\uu + p_{\phi}(\rb,t)\partial_{\phi}\uu \in\TS$ with scalar functions $p_{\theta}(\rb,t)$ and $p_{\phi}(\rb,t)$ and the tangent space $\TS$ of the sphere $\Sp$ in the point $\rb$. 

In the free-energy functionals \eqref{eq:F_psi} and \eqref{eq:F_p} we have to replace the integration over the plane $\R^{2}$ by an integration over the sphere $\Sp$ and the Cartesian Laplace operator $\Laplace$ by the surface Laplace-Beltrami operator $\LaplaceS=\DivS\,\GradS$. 
With these replacements, $\mathcal{F}_{\psi}$ and $\mathcal{F}_{\pb}$ become 
{\allowbreak\begin{align}%
\mathcal{F}_{\psi}[\psi] &= \int_{\Sp}\!\! \Big( \frac{1}{2}\psi\big(\TempR + (1 + \LaplaceS)^{2}\big)\psi + \frac{1}{4}\psi^{4} \Big) \dif^{2}r \,, \label{eq:F_psi_S} \\
\mathcal{F}_{\pb}[\pb] &= \int_{\Sp}\!\! \Big( \frac{1}{2} C_{1} \norm{\pb}^{2} + \frac{1}{4}C_{2} \norm{\pb}^{4} \Big) \dif^{2}r \,, \label{eq:F_p_S}%
\end{align}}%
respectively. Here, 
{\allowbreak\begin{align}%
\GradS\psi &= \frac{1}{R}\Big( (\partial_{\theta}\uu)\partial_{\theta}\psi
+ \frac{1}{\sin(\theta)^{2}}(\partial_{\phi}\uu)\partial_{\phi}\psi \Big) , \label{eq:grad_S} \\
\DivS\pb &= \frac{1}{R}\big( \cot(\theta)p_{\theta} + \partial_{\theta} p_{\theta} + \partial_{\phi} p_{\phi} \big)
\label{eq:div_S}%
\end{align}}%
are the gradient and divergence operators in spherical coordinates, respectively. 
In the dynamic equations \eqref{eq:apfc_planeI} and \eqref{eq:apfc_planeII}, we have to restrict the dynamics to the sphere $\Sp$. For the scalar quantity $\psi$ this has already been done in Refs.\ \cite{Backofen2010,Backofen2011}; for the vector quantity $\pb$ we follow the treatment of a surface polar orientation field in Ref.\ \cite{Nestler2018}.
Therefore, we replace the Cartesian Laplace operator $\Laplace$ acting on the scalar-valued $\delta\mathcal{F}/\delta\psi$ by the surface Laplace-Beltrami operator $\LaplaceS$, the Laplace operator $\Laplace$ acting on the vector-valued $\delta\mathcal{F}/\delta\pb$ by $-\LaplaceDR$ with the surface Laplace-de\,Rham operator 
$\LaplaceDR=-\GradS\DivS-\rotS\RotS$, where
{\allowdisplaybreaks\begin{align}%
\!\!\!\rotS\psi &= \frac{1}{R\sin(\theta)}\big(-(\partial_{\theta}\uu)\partial_{\phi}\psi 
+ (\partial_{\phi}\uu)\partial_{\theta}\psi\big) \,, \label{eq:rot_S} \\
\!\!\!\RotS\pb &= \frac{1}{R}\Big(2\cos(\theta)p_{\phi} \!-\! \frac{1}{\sin(\theta)}\partial_{\phi} p_{\theta} + \sin(\theta)\partial_{\theta} p_{\phi}\Big) \label{eq:Rot_S}%
\end{align}}%
are the surface curl operators in spherical coordinates, as well as $\Grad$ and $\Div$ by $\GradS$ and $\DivS$, respectively.
This results in the dynamic equations
{\allowbreak\begin{align}%
\partial_{t} \psi &= \LaplaceS\frac{\delta\mathcal{F}_{\psi}}{\delta\psi} - v_{0}\,\DivS\pb \,, \label{eq:apfc_SI} \\
\partial_{t} \pb &= -(\LaplaceDR + D_{r})\frac{\delta\mathcal{F}_{\pb}}{\delta\pb} - v_{0}\,\GradS\psi \,, \label{eq:apfc_SII}
\end{align}}%
which describe active-particle transport tangential to $\Sp$. Together with Eqs.\ \eqref{eq:F_psi_S} and \eqref{eq:F_p_S}, the dynamic equations \eqref{eq:apfc_SI} and \eqref{eq:apfc_SII} constitute a minimal field theoretical model for active crystals on a sphere. 
This model is an extension of the previously proposed model \eqref{eq:apfc_planeI}-\eqref{eq:F_p} for the plane and locally reduces to the latter in the limit $R\to\infty$. For $v_{0}=0$, Eq.\ \eqref{eq:apfc_SI} reduces to the traditional PFC model on a sphere, describing crystallization of passive particles on a sphere \cite{Koehler2016}.

\section{\label{sec:numerical_solution}Numerical solution of the PFC model}
In order to study active crystals on a sphere, we solved the PFC equations \eqref{eq:apfc_SI} and \eqref{eq:apfc_SII} numerically. 
For this purpose, we expanded $\psi$ and $\pb$ in (vector) spherical harmonics so that the partial differential equations \eqref{eq:apfc_SI} and \eqref{eq:apfc_SII} reduce to a set of ordinary differential equations for the time-dependent expansion coefficients of $\psi$ and $\pb$. 
In the following, we first address this (vector) spherical harmonics expansion in more detail. 
Afterwards, we describe for which parameters and setups we solved the dynamic equations and how we analyzed the results.

\subsection{\label{sec:spherical_harmonics}(Vector) spherical harmonics expansion}
In order to discretize Eqs.\ \eqref{eq:apfc_SI} and \eqref{eq:apfc_SII} on the sphere, an expansion of the fields $\psi$ and $\pb$ based on spherical harmonics is used.

We start with the scalar field $\psi$. Let $\mathcal{I}_{\Nsh}=\{(l,m)\,:\,0\leq l\leq \Nsh,\,|m|\leq l\}$ be an index set of the spherical harmonics $Y^{m}_{l}:\Sp\to\C$ up to order $\Nsh$. As an orthonormal set of eigenfunctions of the Laplace-Beltrami operator $\LaplaceS$, with
\begin{equation}
\LaplaceS Y_l^m(\rb) = -\frac{l(l+1)}{R^2}Y_l^m(\rb)  \quad\text{for }(l,m)\in \mathcal{I}_\infty \,,
\end{equation}
the spherical harmonics are dense in the function space $L^2(\Sp)$ \cite{Freeden1994,Freeden2009}.
Therefore, the scalar field $\psi$ can be represented as the series expansion
\begin{equation}
\psi(\rb,t) = \!\!\! \sum_{(l,m)\in\,\mathcal{I}_{\infty}} \!\!\! \hat{\psi}_{lm}(t) Y_l^m(\rb)
\end{equation}
with the expansion coefficients $\hat{\psi}_{lm}(t)$. 

Considering the time dependence of $\psi$ and $\pb$ temporarily as a parameter (so that they become functions of only $\rb$) to simplify the notation, we now address the vector field $\pb:\Sp\to\mathrm{T}\Sp$ with the tangent bundle $\mathrm{T}\Sp$ of the sphere $\Sp$. For this vector field, a different expansion than for $\psi$ is needed. 
Since every continuously differentiable spherical tangent vector field $\pb:\Sp\rightarrow \textup{T}\Sp$ can be decomposed into a curl-free field and a divergence-free field \cite{Freeden2009}, there exist differentiable scalar functions $p_{1},p_2\in C^{1}(\Sp)$ with
\begin{equation}
\pb(\rb,t) = \GradS p_{1}(\rb,t) + \rotS p_2(\rb,t) \,.
\end{equation}
Therefore, a tangent vector field basis can be constructed from the gradient $\GradS$ and curl $\rotS=\uu\times\GradS$ of the spherical harmonics basis functions. We introduce the vector spherical harmonics 
{\allowdisplaybreaks\begin{align}%
\mathbf{y}_{lm}^{(1)}(\rb) &= R\, \GradS Y_l^m(\rb) \,, \\
\mathbf{y}_{lm}^{(2)}(\rb) &= -\frac{\rb}{\norm{\rb}}\times \mathbf{y}_{lm}^{(1)}(\rb)
\end{align}}%
that form an orthogonal system of eigenfunctions of the Laplace-de\,Rham operator $\LaplaceDR$ with
\begin{equation}
\LaplaceDR \mathbf{y}_{lm}^{(i)}(\rb) = \frac{l(l+1)}{R^2}\mathbf{y}_{lm}^{(i)}(\rb) 
\end{equation}
for $i\in\{1,2\}$ and $(l,m)\in\mathcal{I}_{\infty}$.
Also these vector basis functions build a dense function system so that a series expansion of $\pb$ in $\mathbf{y}_{lm}^{(i)}$ is possible:
\begin{equation}
\pb(\rb,t)=\sum_{i=1}^2\sum_{(l,m)\in\,\mathcal{I}_{\infty}} \!\!\! \pbc_{lm}^{(i)}(t) \mathbf{y}_{lm}^{(i)}(\rb) \,.
\end{equation}
Here, $\pbc_{lm}^{(i)}(t)$ are the scalar expansion coefficients of $\pb$. 

Introducing the spaces 
{\allowdisplaybreaks\begin{align}%
\Pi^{\psi}_{\Nsh}(\Sp) &= \Big\{ \mathbf{\psi}=\sum_{(l,m)\in\,\mathcal{I}_{\Nsh}} \!\!\! \hat{\psi}_{lm} Y_l^m \Big\} \,,\\
\Pi^{\pb}_{\Nsh}(\Sp) &= \Big\{ \pb=\sum_{i=1}^2\sum_{(l,m)\in\,\mathcal{I}_{\Nsh}} \!\!\! \pbc_{lm}^{(i)} \mathbf{y}_{lm}^{(i)} \Big\} 
\end{align}}%
of truncated (vector) spherical harmonics expansions of $\psi$ and $\pb$, the polar active crystal equations
{\allowdisplaybreaks\begin{align}%
\partial_{t} \psi &= \LaplaceS\big((\TempR + (1+\LaplaceS)^2)\psi + \nu\big) - v_{0}\,\DivS\pb \,, \label{eq:apfc_fullI} \\
\partial_{t} \pb &= -(\LaplaceDR + D_{r})\big(C_{1}\pb + C_{2}\mathbf{q}\big) - v_{0}\,\GradS\psi \label{eq:apfc_fullII}
\end{align}}%
with the nonlinear terms $\nu=\psi^3$ and $\mathbf{q}=\norm{\pb}^2\pb$ can be formulated in terms of a Galerkin method \cite{Hesthaven2007}.
Therefore, we expand $\psi$ and $\nu$ in $\Pi^{\psi}_{\Nsh}(\Sp)$ and $\pb$ and $\qb$ in $\Pi^{\pb}_{\Nsh}(\Sp)$ and require the residual of Eqs.\ \eqref{eq:apfc_fullI} and \eqref{eq:apfc_fullII} to be orthogonal to $\Pi^{\psi}_{\Nsh}(\Sp)\times\Pi^{\pb}_{\Nsh}(\Sp)$. This leads to the Galerkin scheme
{\allowdisplaybreaks\begin{align}%
\begin{split}%
\partial_{t} \hat{\psi}_{lm}(t) &+ \frac{l(l+1)}{R^2}\Big(\TempR+\Big(1-\frac{l(l+1)}{R^2}\Big)^2\Big)\hat{\psi}_{lm}(t) \\
&+ \frac{l(l+1)}{R^2}\hat{\nu}_{lm}(t) - v_{0} \frac{l(l+1)}{R}\pbc_{lm}^{(1)}(t) = 0 \,,
\label{eq:discrete_pfc}%
\end{split}\\
\begin{split}%
\partial_{t} \pbc_{lm}^{(i)}(t) &+ \Big(\frac{l(l+1)}{R^2} + D_r\Big)\big(C_{1}\pbc_{lm}^{(i)}(t) + C_{2}\qbc_{lm}^{(i)}(t)\big) \\
&+ v_{0} \frac{\delta_{i1}}{R}\hat{\psi}_{lm}(t) = 0 
\label{eq:discrete_polar}%
\end{split}%
\end{align}}%
for $i\in\{1,2\}$, $(l,m)\in\mathcal{I}_{\Nsh}$, and $t\in[t_{0},t_{\mathrm{end}}]$, where $\hat{\nu}_{lm}(t)$ are the expansion coefficients of $\nu$, $\qbc_{lm}^{(i)}(t)$ are the expansion coefficients of $\qb$, $\delta_{ij}$ is the Kronecker delta function, and $t_{\mathrm{end}}$ is the length of the simulated time interval starting at $t_{0}=0$. 

The identification of the expansion coefficients $\hat{\psi}_{lm}$ and $\pbc_{lm}^{(i)}$ for given $\psi$ and $\pb$ requires the evaluation of $L^2$ inner products $\langle\psi,\, Y_l^m\rangle_{\Sp}$ and $\langle\pb,\, \mathbf{y}_{lm}^{(i)}\rangle_{\Sp}$ and thus quadrature on the sphere $\Sp$. This is realized by evaluating $\psi$ and $\pb$ in Gaussian points $\{(\theta_{i}, \phi_{j}) \,:\, 1\leq i \leq\Nth, 1\leq j \leq\Nphi\}$, where $\Nth$ and $\Nphi$ are the numbers of grid points along the polar and azimuthal coordinates, respectively, and utilizing an appropriate quadrature rule \cite{Schaeffer2013}.
For the time-discretization of Eqs.\ \eqref{eq:discrete_pfc} and \eqref{eq:discrete_polar}, a second-order accurate scheme similar to that described in Ref.\ \cite{Backofen2011} is applied. Our implementation of the vector spherical harmonics is based on the toolbox SHTns \cite{Schaeffer2013}.

\subsection{\label{sec:Parameters}Parameters and analysis}
When solving the PFC model for active crystals on a sphere numerically, we considered two setups with different simulation parameters (see Tab.\ \ref{tab1}).
\begin{table}[ht]
\begin{ruledtabular}%
\begin{tabular}{lcc}%
\textbf{Parameters} & \textbf{Setup 1} & \textbf{Setup 2} \\\hline
$R$ & $20$ & $80$ \Tstrut \\
$\bar{\psi}$ & $-0.4$ & $-0.4$ \\
$\TempR$ & $-0.98$ & $-0.98$ \\
$C_{1}$ & $0.2$ & $0.2$ \\
$C_{2}$ & $0$ & $0$ \\
$D_{r}$ & $0.5$ & $0.5$ \\
$v_{0}$ & $[0,0.8]$ & $[0,0.8]$ \\
$\Nsh$ & $250$ & $500$ \\
$\Nth$ & $256$ & $512$ \\
$\Nphi$ & $512$ & $1024$ \\
$t_{0}$ & $0$ & $0$ \\
$t_{\mathrm{end}}$ & $5000$ & $5000$ \\
$\tau$ & $0.005$ & $0.005$ 
\end{tabular}%
\end{ruledtabular}%
\caption{\label{tab1}Simulation parameters for the two different setups we considered.}%
\end{table}
In the first one, the sphere has radius $R=20$ and a crystal that covers the sphere consists of approximately $120$ density maxima (``particles''); in the second setup, the sphere has the larger radius $R=80$ leading to a crystal with approximately $1800$ density maxima. The mean value of the field $\psi$ is $\bar{\psi}=-0.4$ in both cases. We used this value to allow a direct comparison of some of our results (see below) with corresponding results for the flat space presented in Ref.\ \cite{MenzelL2013}. For the same reason, we chose always the parameters in Eqs.\ \eqref{eq:F_psi} and \eqref{eq:F_p} as $\TempR=-0.98$, $C_{1}=0.2$, and $C_{2}=0$ and the rescaled rotational diffusion coefficient as $D_r=0.5$. Regarding the activity parameter $v_{0}$, values in the interval $[0,0.8]$ are considered for both setups. This interval turned out to be appropriate for observing the active crystals that constitute the scope of this work. 
The maximal order $\Nsh$ at which the (vector) spherical harmonics expansions described in Sec.\ \ref{sec:spherical_harmonics} are truncated, is chosen as $\Nsh=250$ in the first and $\Nsh=500$ in the second setup. Furthermore, the parameters $\Nth$ and $\Nphi$ defining the resolution of the grid of Gaussian points on the sphere are $\Nth=256$ and $\Nphi=512$ in the first setup and $\Nth=512$ and $\Nphi=1024$ in the second setup. 
All simulations started from a slightly inhomogeneous random initial density field $\psi(\rb,0)$ and a vanishing initial polarization field $\pb(\rb,0)=\mathbf{0}$. We ran the simulations from $t_{0}=0$ to $t_{\mathrm{end}}=5000$ with time-step size $\tau=0.005$. 

\begin{figure}
\begin{center}%
\includegraphics[width=\linewidth]{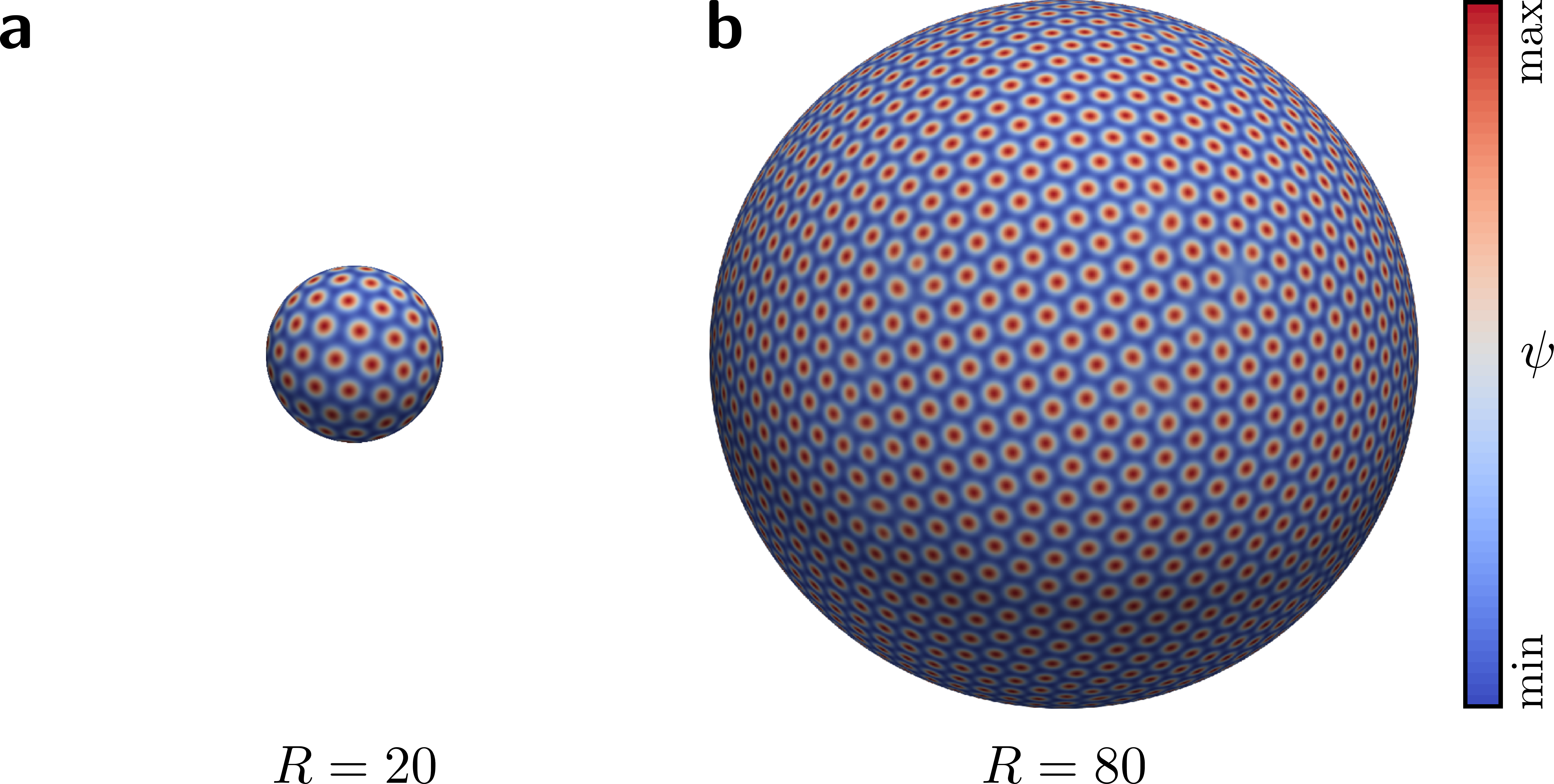}%
\end{center}%
\caption{\label{fig1}(Color online) Late-time density field $\psi(\rb,t)$ on spheres with radii (a) $R=20$ and (b) $R=80$, relaxed from a noisy initial density to a crystalline state containing defects for $v_{0}=0.31$.}
\end{figure}

For the simulation parameters considered in this work, the time-evolution of the density and polarization fields $\psi(\rb,t)$ and $\pb(\rb,t)$, respectively, leads to a crystalline state with local density maxima that can be interpreted as particles forming a crystal (see Fig.\ \ref{fig1} for an example).  
To analyze the emerging patterns of $\psi$ and $\pb$, we introduce some appropriate quantities. 

For characterizing the density field, we identify the positions $\rb_{i}(t)$ with $i\in\{1,\dotsc,n_{p}(t)\}$ of the local density maxima (``particle positions''), where $n_{p}(t)$ is their total number in the considered density field at time $t$. The set $\{\rb_{i}(t)\}$ of the particle positions at time $t$ is defined as 
\begin{equation}
\Big\{ \rb\,:\, \psi(\rb,t) = \max\{\psi(\rb',t)\,:\,\rb' \in \mathcal{U}(\rb)\} \Big\} 
\end{equation}
with $\mathcal{U}(\rb) = \mathcal{B}(\frac{d}{2},\rb)\cap\Sp$ being a neighborhood around $\rb$ on the sphere $\Sp$, 
where $\mathcal{B}(\frac{d}{2},\rb)$ is an open ball of radius $d/2$ centered at $\rb$ and $d=4\pi/\sqrt{3}$ is the center-to-center distance of neighboring particles in a flat hexagonal lattice with lattice constant $2\pi$.  
Since the positions $\rb_{i}(t)$ can be time-dependent, we calculate also their velocities 
\begin{equation}
\vb_{i}(t) = \frac{1}{\tau} \big(\rb_{i}(t) - \rb_{i}(t-\tau)\big) \,.
\end{equation}
Averaging the velocities $\vb_{i}(t)$ locally over an appropriate time interval, which is $3000\leq t\leq 4000$ in this work, and spatial smoothing yields a continuous local velocity field $\vb_{l}(\rb)$ that gives insights into the particle motion at late times.   
The mean particle speed $\vm$ in the crystalline state is obtained as 
\begin{equation}
\vm = \frac{1}{t_{\mathrm{end}} - t_{c}} \int^{t_{\mathrm{end}}}_{t_{c}} \!\! 
\frac{1}{n_{p}(t)} \sum^{n_{p}(t)}_{i=1}  \norm{\vb_{i}(t)} \,\dif t \,,
\end{equation}
where $t_{c}$, which we chose as $t_{c}=1000$, is a sufficiently large time after which the crystalline state has formed.

To characterize the polarization field, we assign a net polarization to each density peak.
For the $i$th particle, being at position $\rb_{i}(t)$, the net polarization $\pb_{i}(t)$ is calculated as 
{\allowdisplaybreaks\begin{align}%
\pb_{i}(t) &= \frac{\tilde{\pb}_{i}(t)}{\norm{\tilde{\pb}_{i}(t)}} \,, \\
\tilde{\pb}_{i}(t) &=  \int_{\mathcal{U}(\rb_{i}(t))} \!\!\!\!\!\!\!\!\!\!\!\! \psi^{+}(\rb,t)\, \pi_{\Tang{i}{\Sp}}(t) \pb(\rb,t)\,\dif^{2}r 
\end{align}}%
with the shifted density field $\psi^{+}(\rb,t)=\psi(\rb,t)-\min_{\rb}(\psi(\rb,t))$, where $\min_{\rb}(\psi(\rb,t))$ is the minimal value of $\psi$ at time $t$, and the projection $\pi_{\Tang{i}{\Sp}}(t)$ that maps onto the tangent plane $\Tang{\rb_{i}(t)}{\Sp}$.
We also define a coarse-grained polar order parameter 
\begin{equation}
\pp_{i}(t) = \frac{1}{\bar{w}_{i}(t)}\sum_{\begin{subarray}{c}j\in\mathcal{J}_{t}(\rb_{i}(t))\\j\neq i \end{subarray}} \!\!\! w_{ij}(t)\, \pb_{i}(t) \bigcdot \pb_{j}(t) \,, 
\label{eq:Pcal_i}%
\end{equation}
which measures the parallelity of the net polarization $\pb_{i}(t)$ of the $i$th particle with respect to the net polarizations of the neighboring particles. 
In Eq.\ \eqref{eq:Pcal_i}, $\mathcal{J}_{t}(\rb) = \{j\,:\, \norm{\rb_{j}(t)-\rb} < r_{\mathrm{cut}}\}$ 
is the index set of the particles with a distance smaller than the cutoff radius $r_{\mathrm{cut}}=2.5d$ from $\rb$ at time $t$.
The weights $w_{ij}(t)$ are chosen as the inverse distance of the $i$th and $j$th particle at time $t$, i.e., 
$w_{ij}(t)=1/\norm{\rb_{j}(t)-\rb_{i}(t)}$, and $\bar{w}_{i}(t)$ is the normalization factor 
\begin{equation}
\bar{w}_{i}(t) = \sum_{\begin{subarray}{c}j\in\mathcal{J}_{t}(\rb_{i}(t))\\j\neq i \end{subarray}} \!\!\! w_{ij}(t) \,.
\end{equation}
By spatially smoothing the discrete polar order parameters $\pp_{i}(t)$ with $i\in\{1,\dotsc,n_{p}(t)\}$, a continuous local  polar order parameter $\pp(\rb,t)$ is obtained.
In addition, we introduce the global polar order parameter
\begin{equation}
\PP(t) = \frac{1}{n_{p}(t)} \sum^{n_{p}(t)}_{i=1} \pp_{i}(t) \,,
\end{equation}
which is a measure for the local parallelity of the particles' net polarizations averaged over the full sphere $\Sp$, 
and the global net polarization vector 
\begin{equation}
\mathbf{P}(t) = \sum^{n_{p}(t)}_{i=1} \pb_{i}(t) \,, 
\end{equation}
which describes the global net polarization of the active crystal.

\section{\label{sec:results}Results}
When calculating the time evolution of $\psi$ and $\pb$ for small $v_{0}$, a crystalline structure builds up (see Fig.\ \ref{fig1}). The polarization field $\pb$ then evolves to nearly the negative gradient direction of $\psi$ forming asters at the density maxima. 

\begin{figure}[ht]
\includegraphics[width=\linewidth]{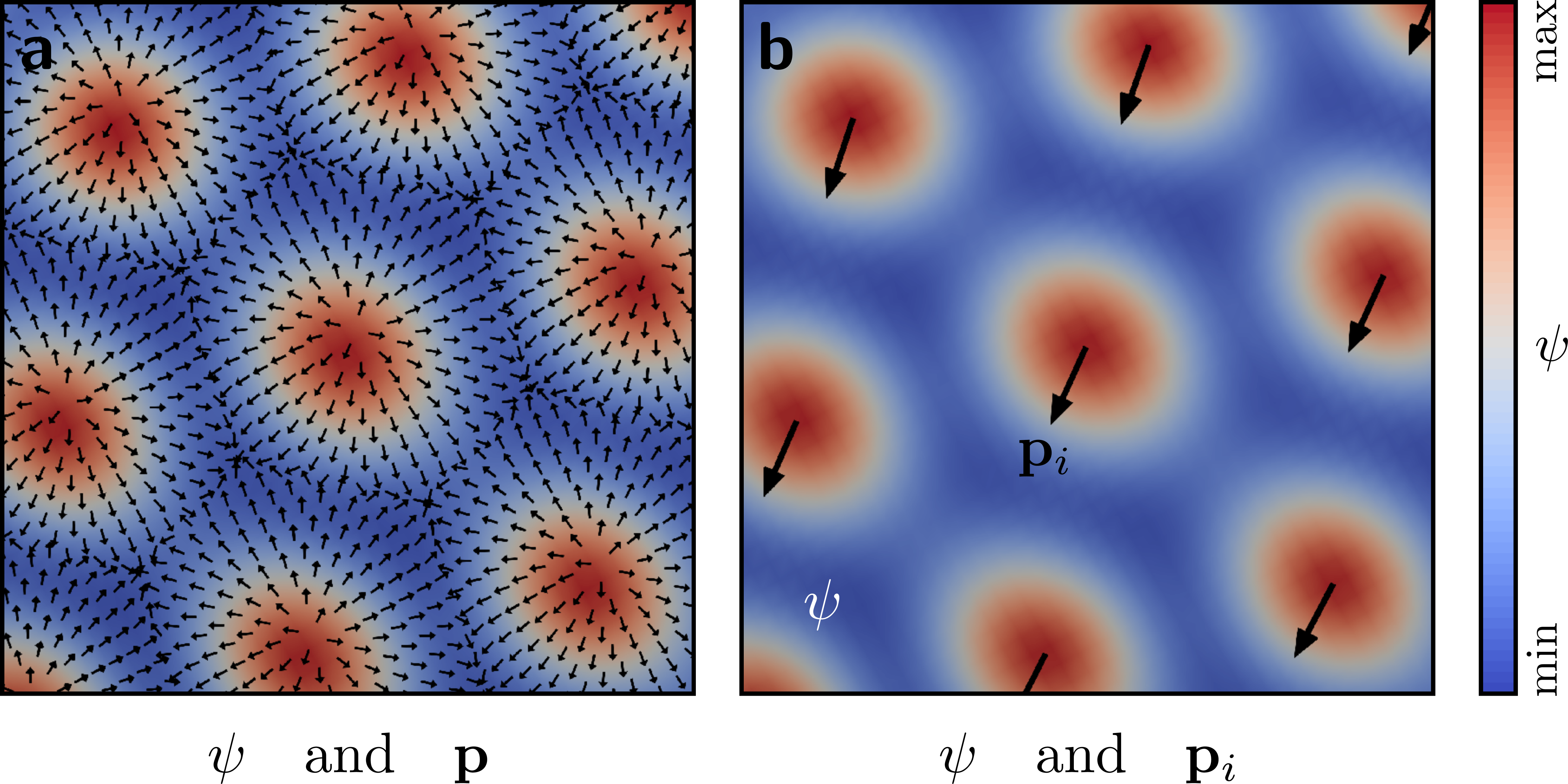}%
\caption{\label{fig2}(Color online) Detailed view on the density field $\psi(\rb,t)$ (background color) from Fig.\ \ref{fig1}b as well as the associated (a) local polarization $\pb(\rb,t)$ (arrows) and (b) normalized net polarizations $\pb_{i}(t)$ (arrows at positions $\rb_{i}$). The crystal moves in the direction of the latter arrows.}
\end{figure}

At a certain threshold value $v_{\mathrm{th}}$ of the activity $v_{0}$, the aster-defect positions of $\pb$ start to depart more and more from the density maxima at $\rb_{i}$ (see Fig.\ \ref{fig2}a), 
leading to nonvanishing net polarizations $\pb_{i}$ (see Fig.\ \ref{fig2}b) and to an advection of the density field.
This means that below this threshold the crystal is static, whereas above the threshold the particles in the crystal move in directions that align with the particles' net polarizations. A similar behavior has been found in the case of a flat periodic domain in Refs.\ \cite{MenzelL2013,Menzel2014,Alaimo2016}. 
For the sphere radii $R$ considered in this work, the activity threshold $v_{\mathrm{th}}$ of the resting to motion transition is $v_{\mathrm{th}}\approx 0.3$, which is smaller than the threshold given in Ref.\ \cite{Menzel2014} for a flat system. 
Both the value for $v_{\mathrm{th}}$ observed in our simulations as well as its apparent independence from $R$ are in very good agreement with results obtained by a linear stability analysis of Eqs.\ \eqref{eq:apfc_SI} and \eqref{eq:apfc_SII} (see Appendix \ref{sec:appendix}). This stability analysis shows that $v_{\mathrm{th}}$ has in fact a nonvanishing but only weak dependence on $R$. The values of $v_{\mathrm{th}}$ vary between a minimum $\vthmin\approx 0.28$ and slightly larger values, where the deviations from $\vthmin$ decrease with growing $R$. For the radii $R=20$ and $R=80$ considered in our simulations, the activity threshold is $v_{\mathrm{th}}\approx 0.31$ and $v_{\mathrm{th}}\approx 0.29$, respectively, and it asymptotically gets constant for $R\to\infty$.  

\begin{figure}[ht]
\includegraphics[width=\linewidth]{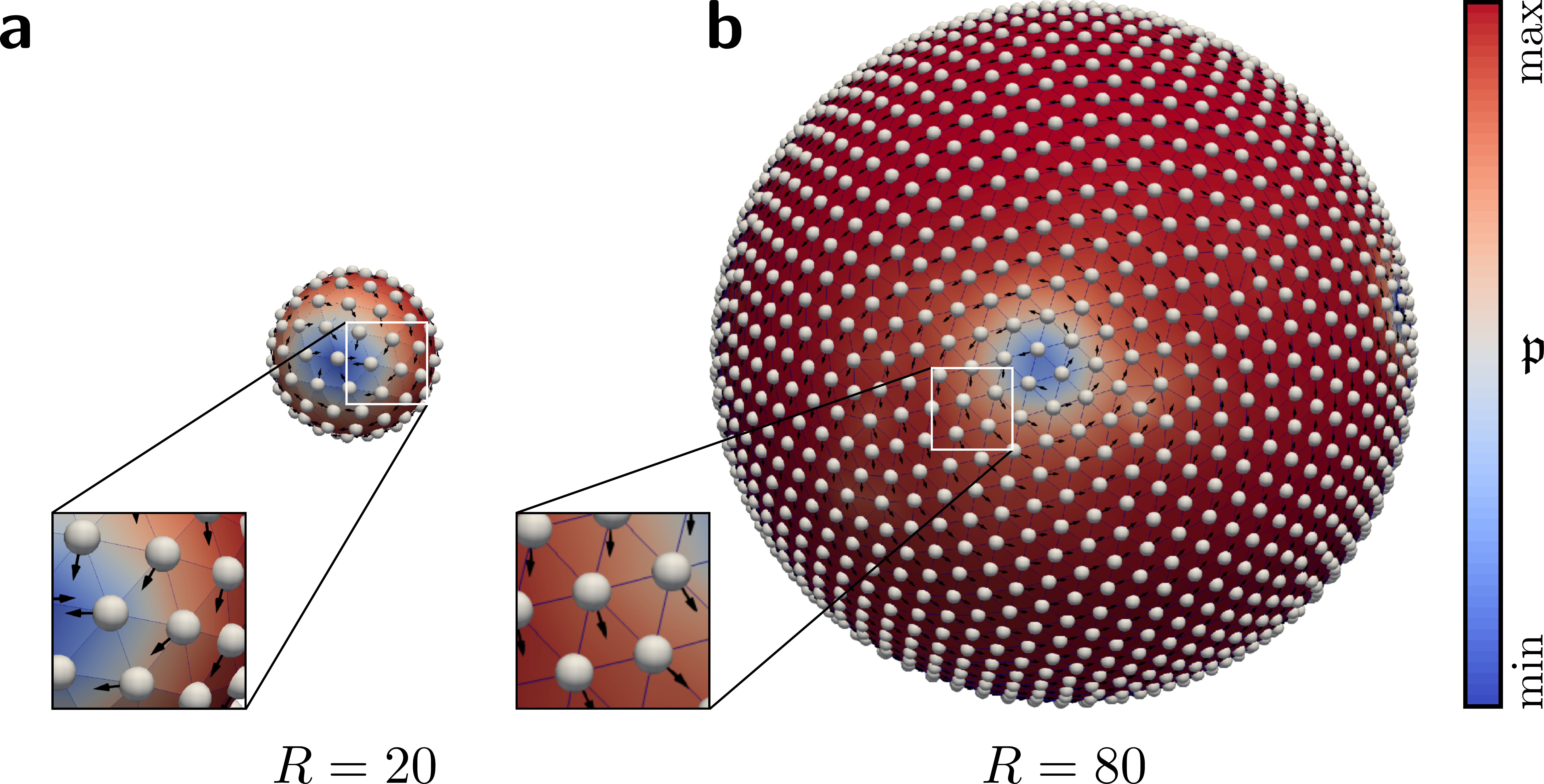}%
\caption{\label{fig3}(Color online) Maxima (visualized as spherical particles) of the density field $\psi$ from Fig.\ \ref{fig1} as well as the associated normalized net polarizations $\pb_{i}$ (arrows, visible in the insets) and local polar order parameter $\pp$ (coloring of the sphere). The crystal has two opposing minima of $\pp$ and rotates about an axis through these minima.}
\end{figure}

For activities not too far above the threshold value, the motion of the individual particles leads to a global motion pattern with a vortex-vortex configuration as shown in Fig.\ \ref{fig3}. 
In this configuration, the net polarizations $\pb_{i}$ form two vortices at oppositely located poles on the sphere, resulting in a self-spinning motion of the crystal about an axis through these poles. Most of the particles in such a self-spinning crystal show a strong parallel local alignment of their net polarizations. The local polar order parameter $\pp(\rb,t)$ of a vortex-vortex crystal has minima at the two poles and it is maximal at the equator.

\begin{figure}[ht]
\includegraphics[width=\linewidth]{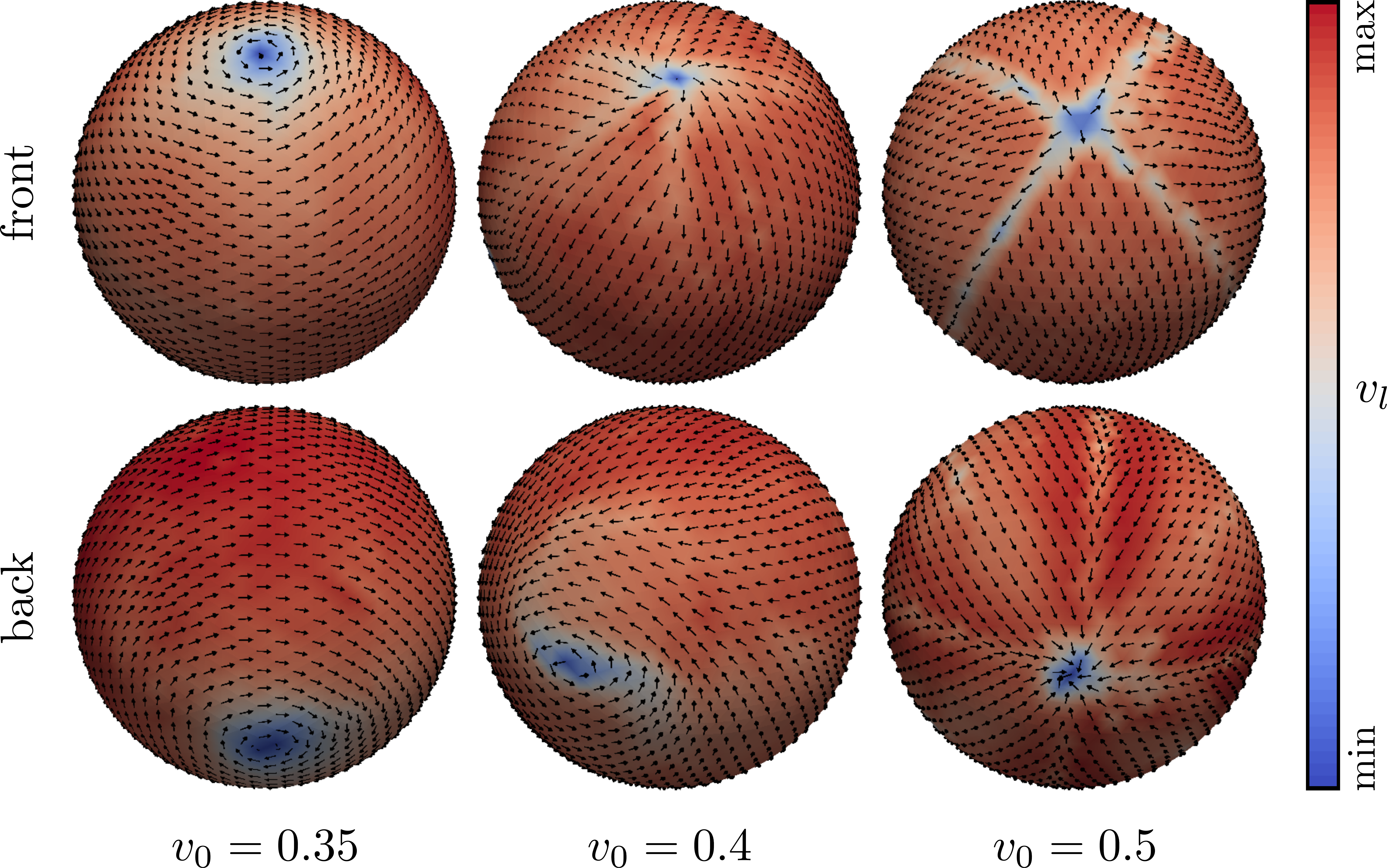}%
\caption{\label{fig4}(Color online) Individual particle velocities $\vb_{i}$ on a sphere with radius $R=80$ averaged over the time interval $3000\leq t\leq 4000$. The coloring of a sphere shows the time-averaged local particle speed $v_{l}(\rb)=\norm{\vb_{l}(\rb)}$ and the arrows show the time-averaged local direction of particle motion $\vb_{l}(\rb)/v_{l}(\rb)$.}
\end{figure}

Figure \ref{fig4} shows the time-averaged local particle velocity $\vb_{l}(\rb)$ for three values of the activity parameter $v_{0}$. 
With increasing $v_{0}$ a transition from a vortex-vortex crystal (left column) to a source-sink crystal (right column) can be seen. 
This transition is smooth, with combinations of vortex and source or sink defects as intermediate states (middle column), and leads to a change in the qualitative behavior of the system. 
While the vortex-vortex crystal seems natural and can be observed directly also by classical particle simulations \cite{Sknepnek2015,JanssenKL2017}, the source-sink crystal, though natural for vector fields \cite{Bowick2009,Sknepnek2015}, must be interpreted in the sense that at one pole the system crystallizes, whereas at the other pole it melts. The form of Eq.\ \eqref{eq:apfc_SI} guarantees mass conservation, but not particle number conservation, which would be expected in a classical discrete particle model.

\begin{figure}[ht]
\includegraphics[width=\linewidth]{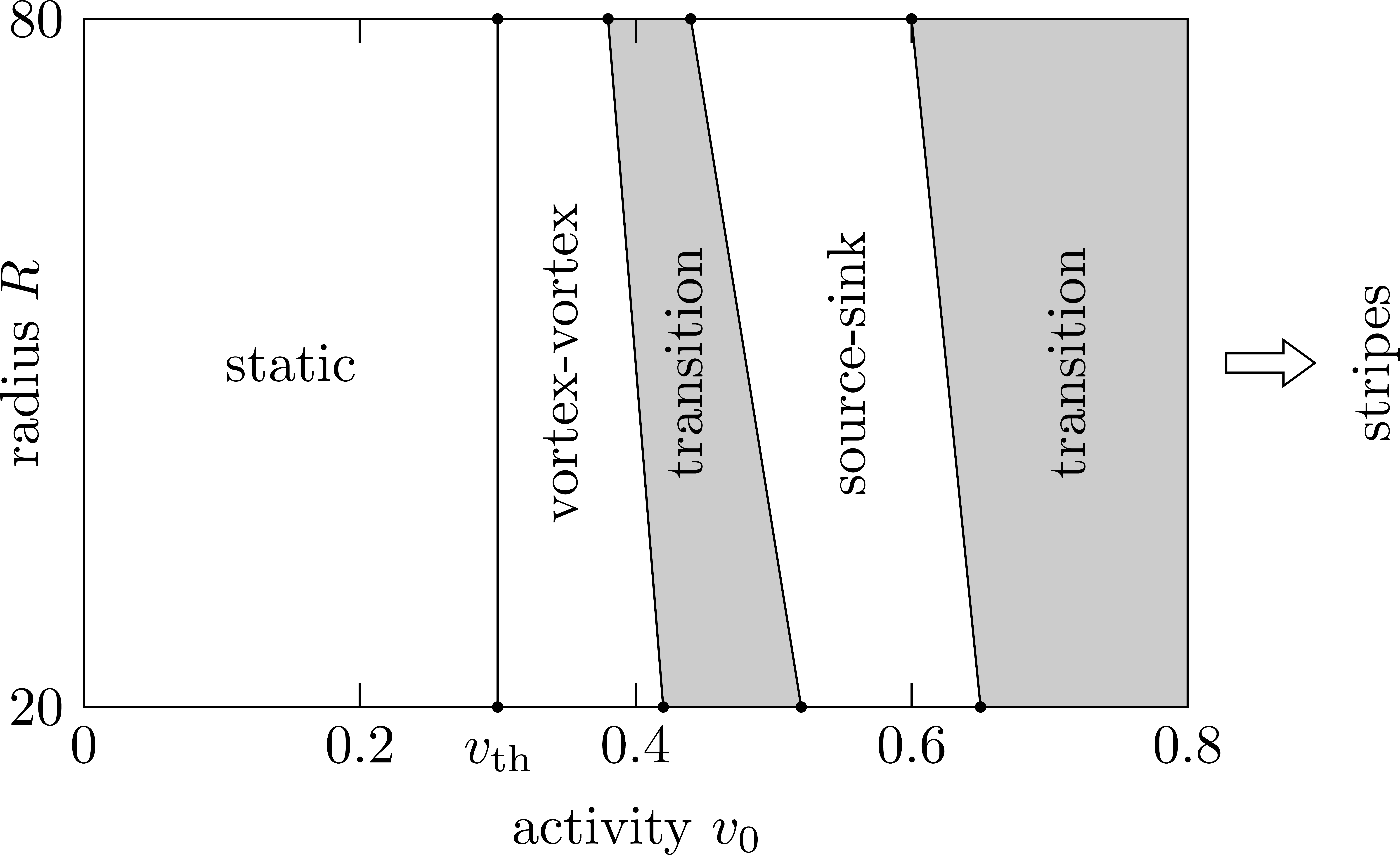}%
\caption{\label{fig5}State diagram for different activities $v_{0}$ and sphere radii $R$ including static, vortex-vortex, and source-sink patterns as well as two broad transition regions. For activities much larger than $v_{0}=0.8$, a stripe state is found.}
\end{figure}

Classification of the observed late-time structures for different $v_{0}$ and $R$ into static, vortex-vortex, and source-sink patterns leads to the state diagram in Fig.\ \ref{fig5}. 
Except for the static to vortex-vortex transition, there are broad transition areas between neighboring states. 
In the transition area between the vortex-vortex and source-sink states, combinations of vortex and source or sink defects are found.
When increasing the activity above $v_{0}=0.8$, a slow transition to a stripe (or lamellar) state is observed. This state is found also in the case of a plain system \cite{MenzelL2013} and for other values of the model parameter $\TempR$ \cite{Achim2011,Stoop2015}. 
For a rather high activity, we found traveling stripes. A more detailed study of this state is, however, beyond the scope of this work.

\begin{figure}[ht]
\includegraphics[width=\linewidth]{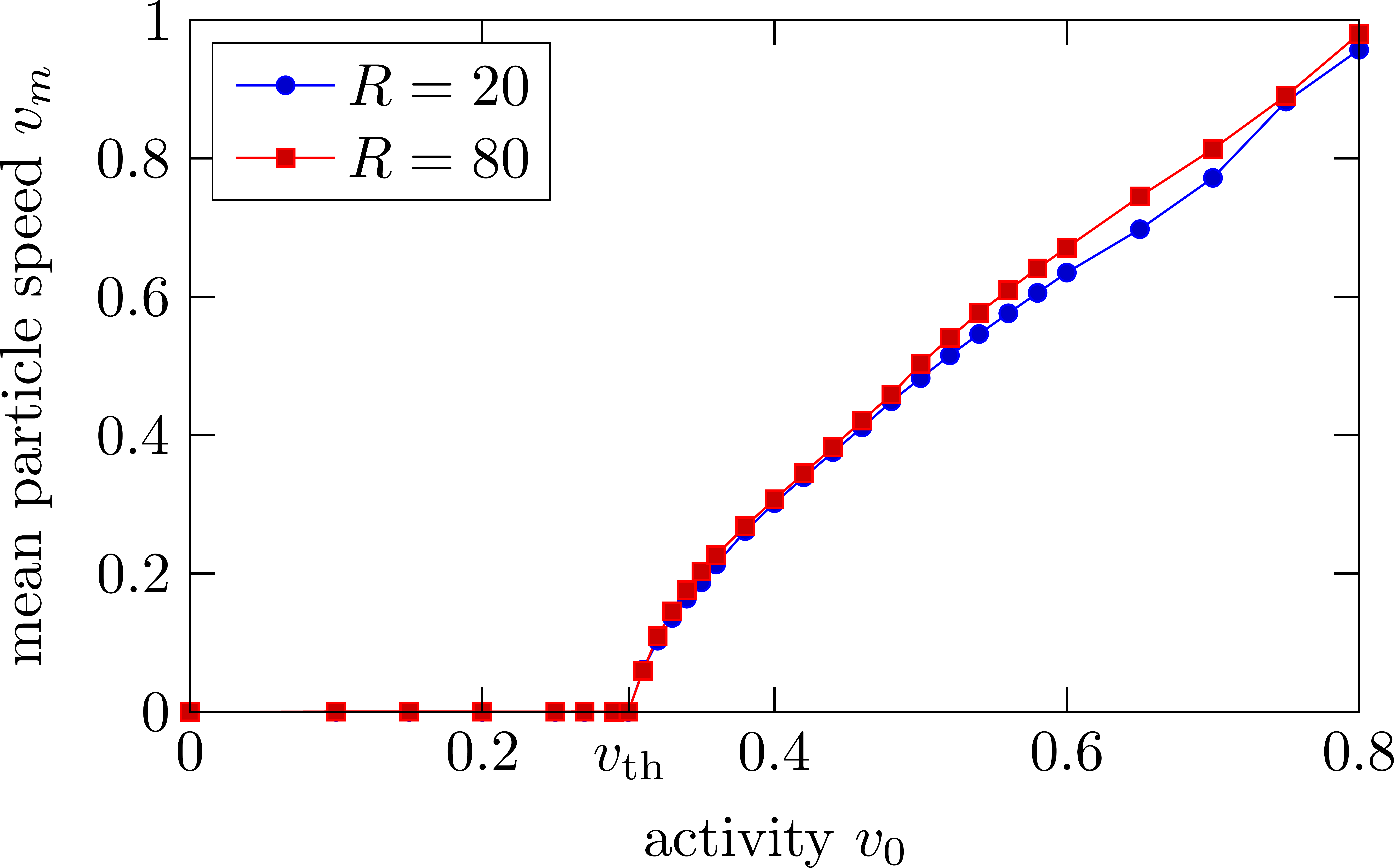}%
\caption{\label{fig6}(Color online) Mean particle speed $\vm$ as a function of activity $v_{0}$ for spheres with radii $R=20$ and $R=80$. For both radii, the activity threshold, where the particles start to move, is $v_{\mathrm{th}}\approx 0.3$. The data are averaged over $50$ simulations.}
\end{figure}

The dependency of the particles' velocities on the activity parameter $v_{0}$ can be studied on the basis of the mean particle speed $\vm$. In Fig.\ \ref{fig6}, it is compared to the activity parameter. 
Interestingly, the function $\vm(v_{0})$ seems to be independent of the sphere radius $R$. The observed behavior is qualitatively the same as  in Refs.\ \cite{Menzel2014,Alaimo2016} for the flat periodic case. However, the absolute value for $v_{\text{th}}$ and the slope for $\vm(v_0)$ differ. 

\begin{figure}[ht]
\includegraphics[width=\linewidth]{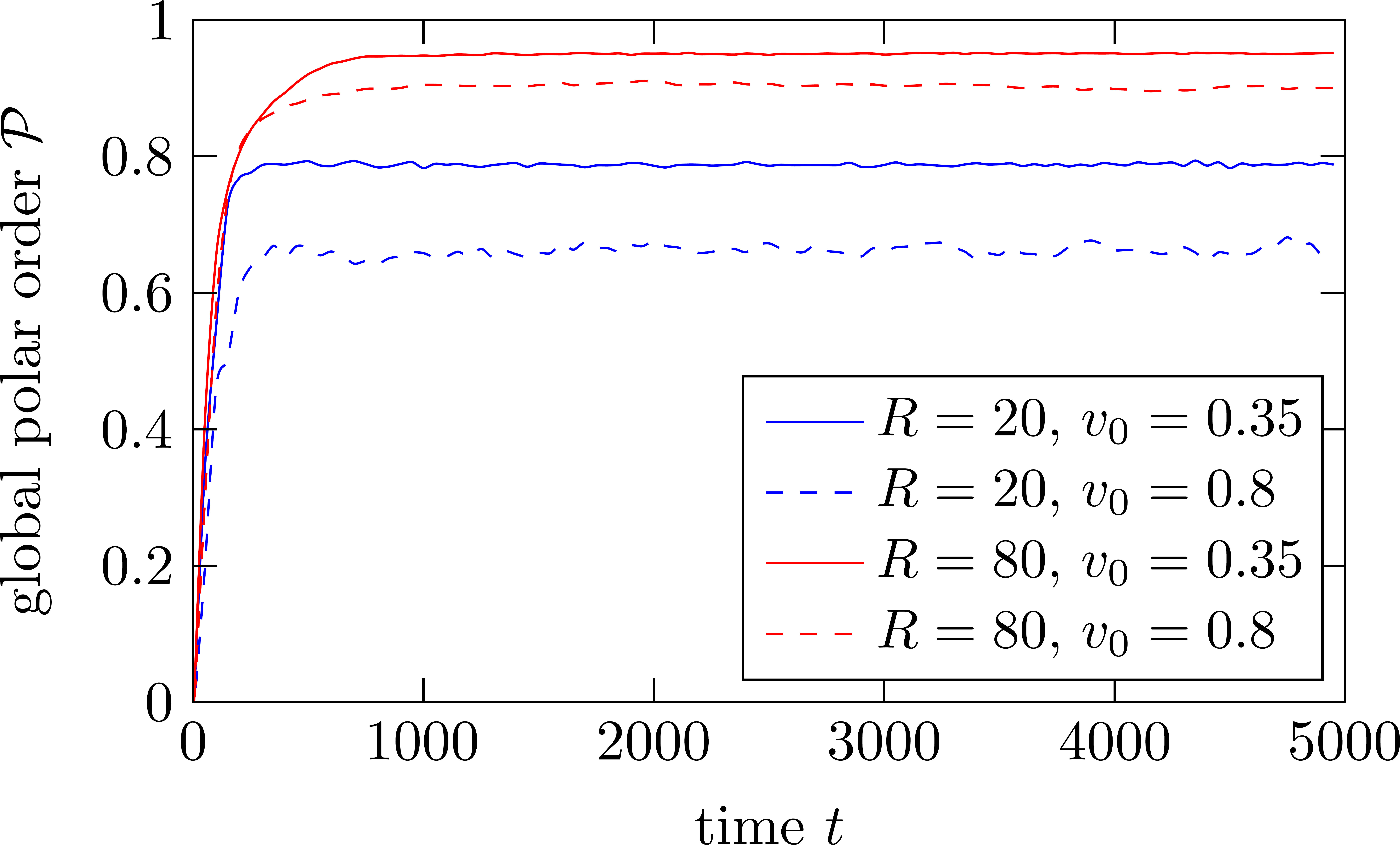}%
\caption{\label{fig7}(Color online) Global polar order parameter $\PP$ as a function of time $t$ for sphere radii $R=20$ and $R=80$ and activities $v_{0}=0.35$ and $v_{0}=0.8$. The data are averaged over $50$ simulations.}
\end{figure}

The net polarizations $\pb_{i}$ build a vector field with global polar order $\PP$. In Fig.\ \ref{fig7}, the polar order parameter $\PP(t)$ is visualized over a large simulation time interval for two radii $R$ of the sphere and two activity parameters $v_{0}$. 
After an initial relaxation at $0\leq t\lesssim 1000$, the global polar order has reached its maximum and stays constant. While for $R=80$ nearly the value $\PP=1$ is reached, a smaller sphere results in a smaller maximal global polar order. This can be explained by the larger geometrical constraints on a sphere with smaller surface area. Also a larger $v_{0}$ leads to a smaller maximal polar order, since for larger $v_{0}$ the particles are more dynamic, which hampers a parallel alignment of the net polarizations.

\begin{figure}[ht]
\includegraphics[width=\linewidth]{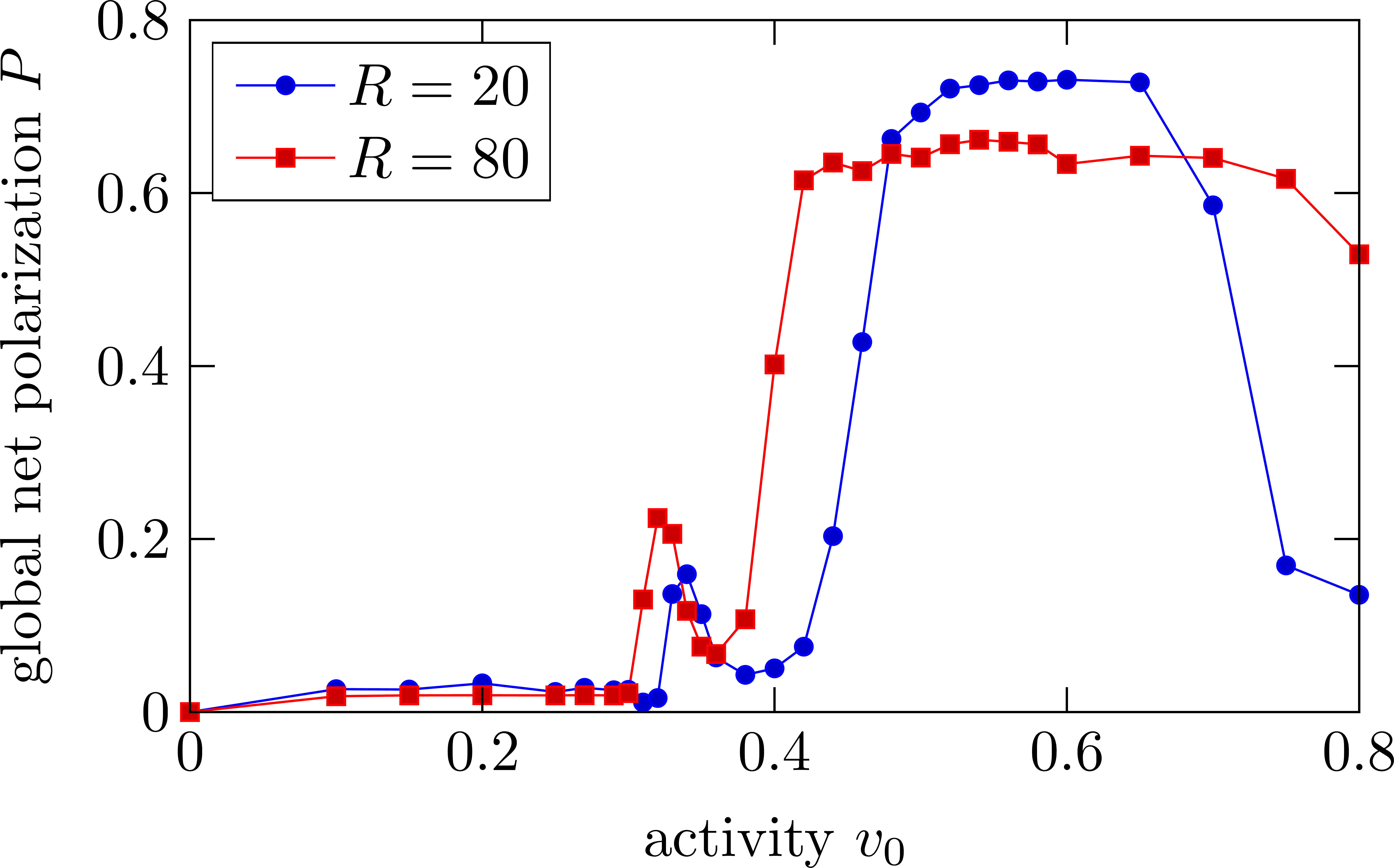}%
\caption{\label{fig8}(Color online) Global net polarization $P=\lVert\mathbf{P}\rVert$ averaged over the time interval $1000\leq t\leq 5000$ as a function of activity $v_{0}$ for spheres with radii $R=20$ and $R=80$. The data are averaged over $50$ simulations.}
\end{figure}

The different states go along also with different values of the time-averaged global net polarization $P=\norm{\mathbf{P}}$ (see Fig.\ \ref{fig8}). 
In the static crystal at small $v_{0}$, the net polarizations $\pb_i$ and thus the global net polarization $P$ vanish. At $v_{0}\approx 0.3$, where nonvanishing net polarizations are established and the density maxima start to move, $P$ suddenly grows; at $v_{0}\approx 0.35$ the global net polarization reaches its next minimum, which is associated with the vortex-vortex crystal; at $v_{0}\approx 0.4$ the global net polarization increases steeply until it reaches its maximum in the source-sink state. The increase of $P$ from the vortex-vortex state to the source-sink state can also be expected from the arrow fields shown in Fig.\ \ref{fig4}. For large activities $v_{0}>0.6$ the source-sink crystal gets disturbed and $P$ decreases again. 

\begin{figure}[ht]
\includegraphics[width=\linewidth]{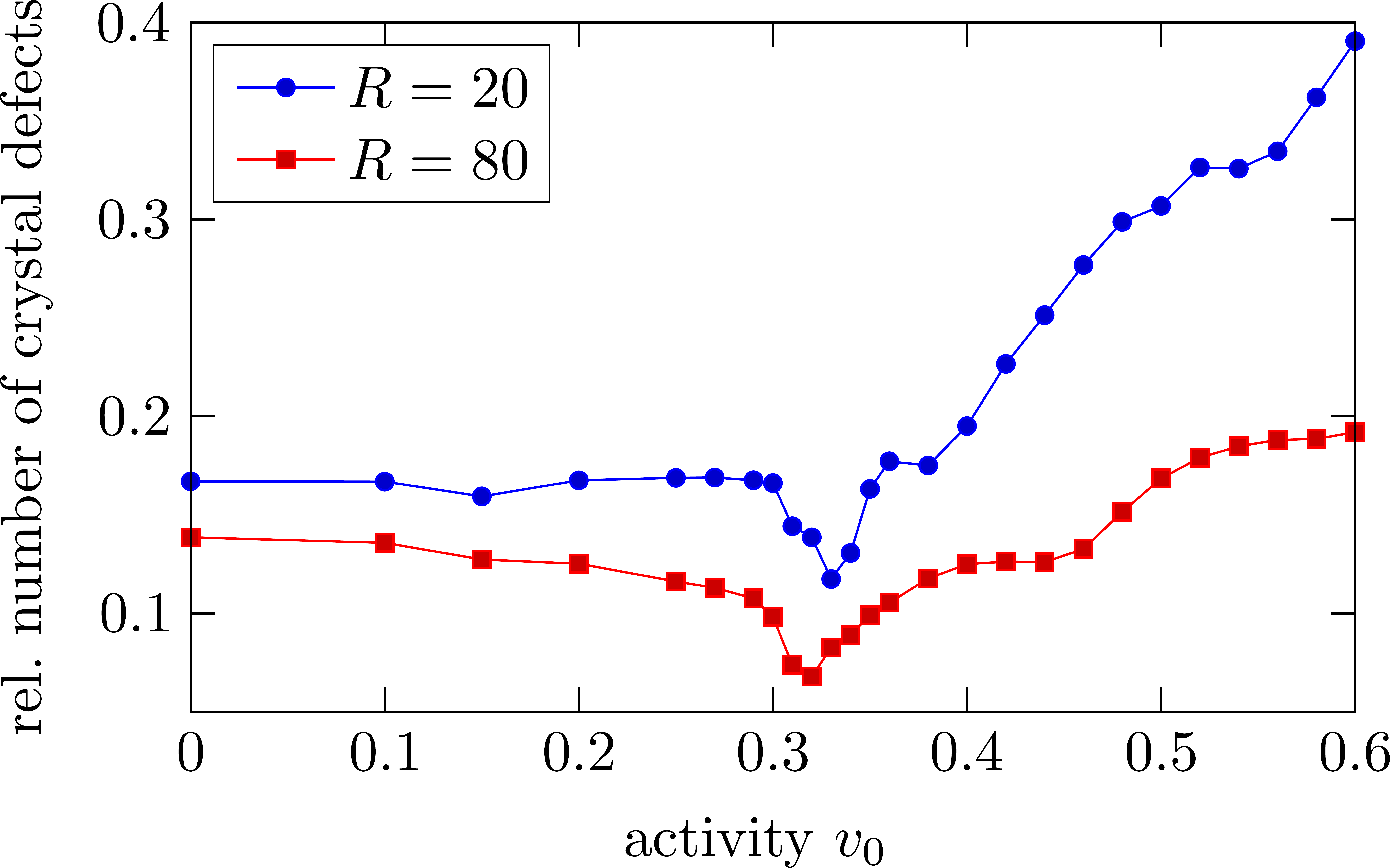}%
\caption{\label{fig9}(Color online) Number of defects in the crystalline structure relative to the overall number of particles as a function of activity $v_{0}$ for spheres with radii $R=20$ and $R=80$. The data are averaged over the time interval $1000\leq t\leq 5000$ and over $50$ simulations.}
\end{figure}

We now focus on the occurrence of translational defects (i.e., dislocations) in the crystalline states, which result from the topological constraints \cite{Bausch2003}. Examples for such defects are already visible in Fig.\ \ref{fig1}. Translational defects in the crystal structure can be identified by the coordination number $\cn_{i}$, which is equal to the number of nearest neighbors of a cell around the node $\rb_{i}$ in a spherical Voronoi diagram with the particle positions $\{\rb_j\}$ as nodes. In a defect-free hexagonal crystal one has $\cn_{i}=6$ for all $i\in\{1,\dotsc,n_{p}\}$, but on a sphere a classical theorem of Euler states that
\begin{equation}
\sum_{\cn} (6-\cn) \nn_{\cn} = 6\chi(\Sp) = 12 \,,
\end{equation}
where $\nn_{\cn}$ is the number of nodes with coordination number $\cn$ and $\chi(\Sp)=2$ is the Euler characteristic of the sphere. Typically, there are no fourfold or lower-order defects in such a crystal. Then the number of fivefold defects is at least $12$ and increases with the number and order of sevenfold and higher-order defects. Counting the total number of defects, i.e., the number of index values $i$ where $\cn_{i}\neq 6$, shows that in our simulations between 10 and 20 percent of the particles in the static crystal ($v_{0} < 0.3$) have more or less than $6$ neighbors (see Fig.\ \ref{fig9}). 
When the activity forces the crystal to move and the vortex-vortex state emerges ($v_{0}\approx 0.3$), the particles are able to improve their spatial arrangement and the number of defects goes down. For larger activities the number of defects increases steeply and it becomes maximal in the source-sink state. This is consistent with the plots in Figs.\ \ref{fig1} and \ref{fig4}, which also indicate that the source-sink crystal contains more defects than the other crystalline states.

\begin{figure}[ht]
\includegraphics[width=\linewidth]{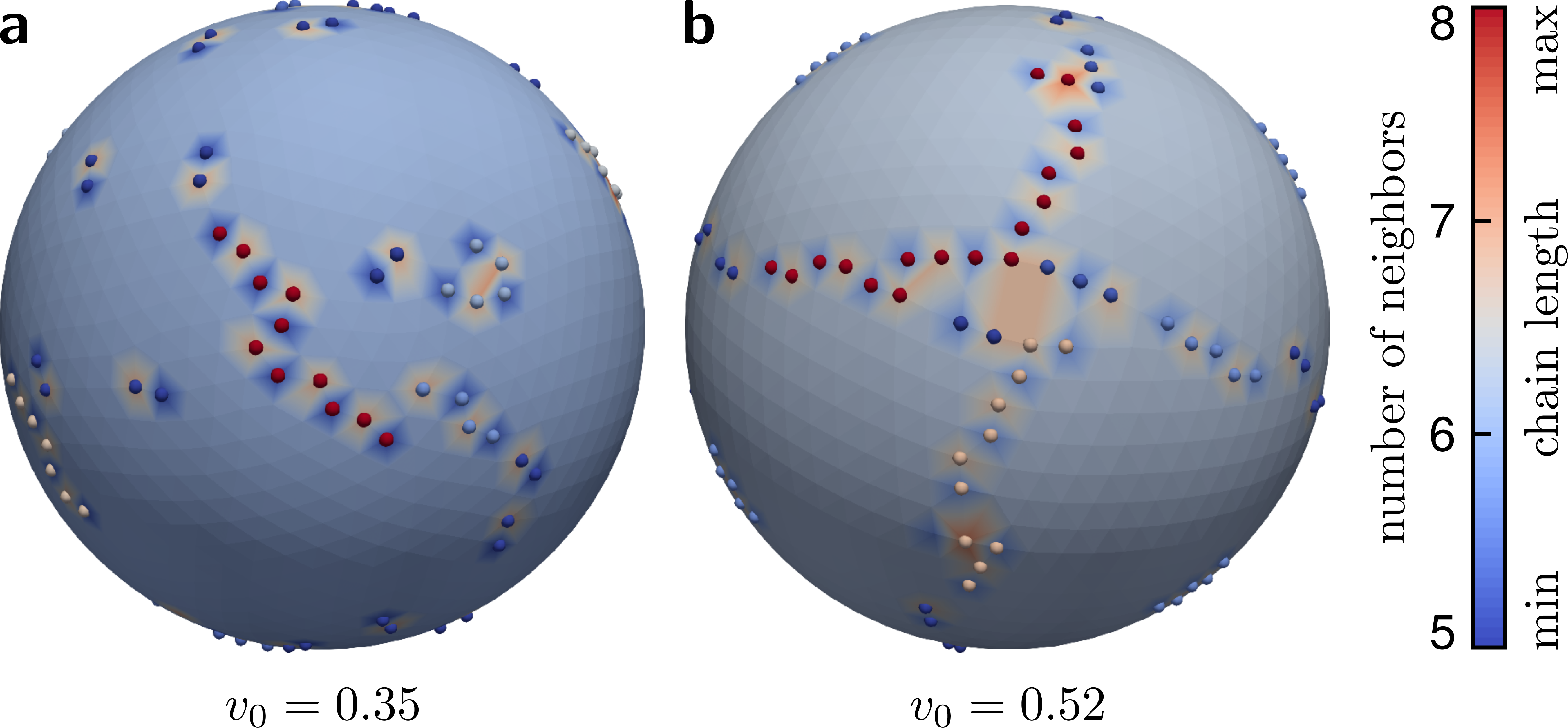}%
\caption{\label{fig10}(Color online) Chains of defects on a sphere with radius $R=80$ at late times for activities (a) $v_{0}=0.35$ and (b) $v_{0}=0.52$. Defects are visualized as small spheres, whose colors indicate the length of the corresponding defect chain. The coloring of the large sphere near a defect depicts the number of maxima of the density field $\psi$ neighboring the defect, where dark blue means 5 neighbors and orange denotes 7 neighbors. Light blue regions of the large sphere are free of defects.}
\end{figure}

\begin{figure}[ht]
\includegraphics[width=\linewidth]{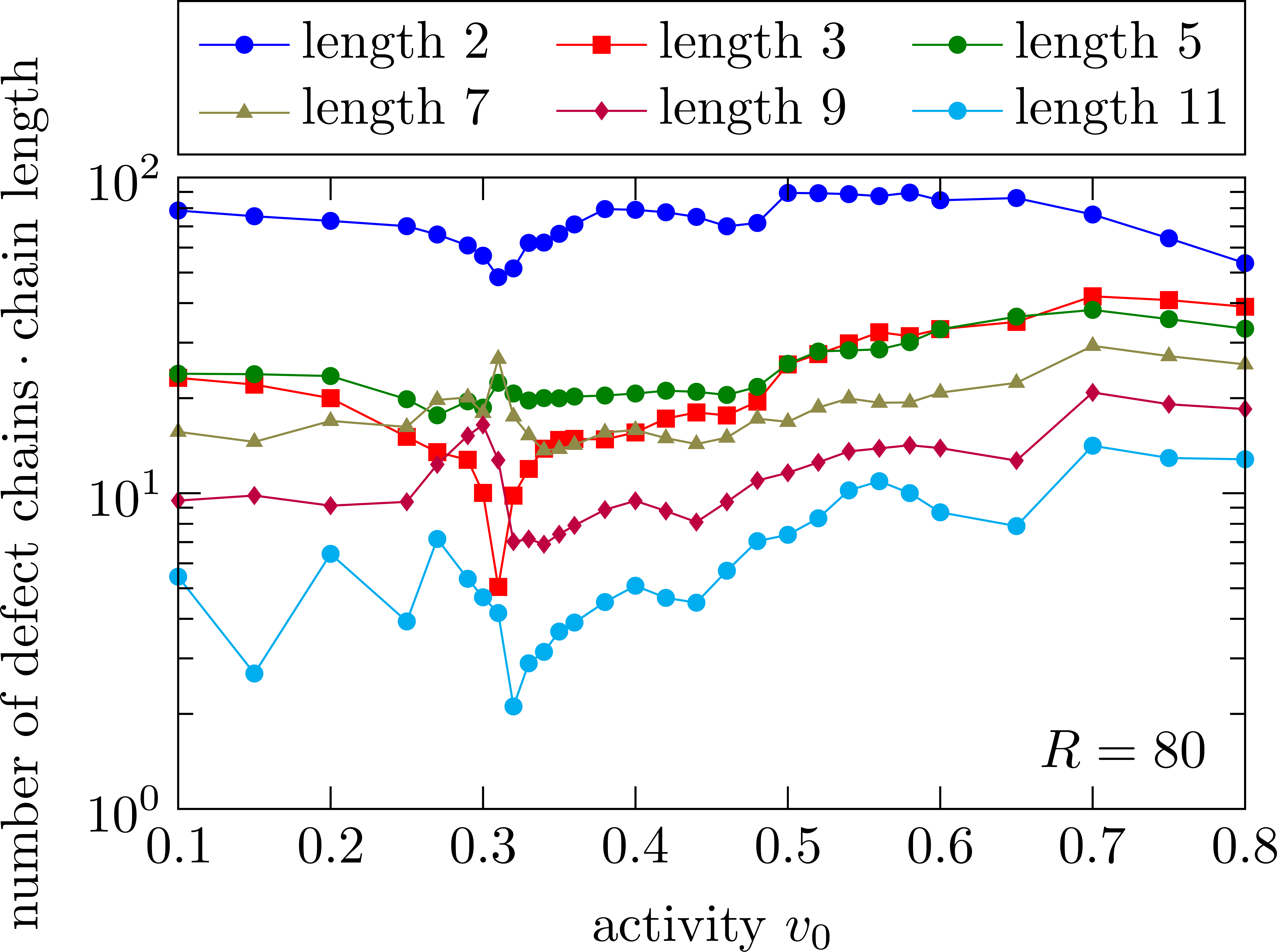}%
\caption{\label{fig11}(Color online) The number of defect chains of a particular length on a sphere with radius $R=80$ multiplied by the chain length, i.e., the total number of defects that are part of the defect chains of this length, as a function of activity $v_{0}$ for six different chain lengths. In the activity range $0.3\lesssim v_{0}\lesssim 0.32$, the numbers of chains with lengths $5$, $7$, and $9$ increase, whereas the numbers of chains with the other lengths decrease. The data are averaged over the time interval $1000\leq t\leq 5000$ and over $50$ simulations.}
\end{figure}

To study the defects in more detail, we now look at chains formed by defects of different coordination numbers (e.g., pairs of five-fold and seven-fold defects). In Fig.\ \ref{fig10}, only the particles with coordination numbers $\cn_{i}\neq 6$ are shown. 
There are separated pairs of defects, but also long chains of defects. This is similar to the passive case ($v_{0}=0$) for various geometries \cite{Bausch2003,Irvine2010,Bendito2013,Schmid2014}, but here the chains of defects are dynamic. Due to the activity of the particles, the defect chains permanently emerge, move, change size, and vanish. 
The numbers of defects located in these chains are statistically analyzed in Fig.\ \ref{fig11} for various activities $v_{0}$. 
Already for small activities there are defect chains of all considered lengths present. Short chains consisting of only two defects are most frequent, whereas with growing length the chains become increasingly rare. This is an overall trend and true for most activities. An exception constitute activities near the threshold value $v_{\mathrm{th}}\approx 0.3$, where the overall number of defects in the system is minimal. 
For such activities, the number of defect chains as a function of the activity $v_{0}$ has an extremum for all chain lengths. Especially very short chains with lengths $2$ and $3$ are less frequent for $v_{0}\approx v_{\mathrm{th}}$ than for smaller or larger values of $v_{0}$. In contrast, the numbers of longer chains with lengths $5$, $7$, and $9$ have a maximum near the threshold activity $v_{\mathrm{th}}$. This means that the vortex-vortex state favors the formation of these longer defect chains.
For larger activities, where the overall number of defects grows with $v_{0}$, the number of defect pairs first increases but later decreases again, whereas the chains consisting of more than two defects increase in number.

\section{\label{sec:conclusions}Conclusions and outlook}
Using a new active phase-field-crystal-type model we have studied crystals of self-propelled colloidal particles on a sphere. 
These ``active crystals'' have a hexagonal local density pattern and -- due to the topological constraints prescribed by the sphere -- always some defects. 
Three types of crystals are observed: a static crystal, a vortex-vortex crystal, and a source-sink crystal. 
When relaxing the particle density field from a random initial density distribution, the number of defects at low activity is $10$-$20$ percent of the total particle number and can be minimized by choosing an activity that corresponds to the vortex-vortex state. 
It should be possible to confirm the observed crystalline states and the results related to their defects by particle-resolved simulations and experiments. 

With the numerical tools for vector-valued surface partial differential equations developed in Refs.\ \cite{Witkowski2015,Nestler2018,Reuther2018}, the problem can even be considered on nonspherical geometries. It would also be interesting to use PFC models to study nonspherical self-propelled particles and their active liquid-crystalline states \cite{SanchezCDHD2012,WensinkDHDGLY2012} on a sphere \cite{JanssenKL2017} and other manifolds \cite{NestlerARXIV2017,SittaSWL2018}. Appropriate PFC models could be obtained by extending the existing PFC models for liquid crystals \cite{WittkowskiLB2010,WittkowskiLB2011,WittkowskiLB2011b,Achim2011,PraetoriusVWL2013} towards active particles and curved manifolds.

\begin{acknowledgments}
We thank Andreas M.\ Menzel and Ingo Nitschke for helpful discussions. 
A.V., R.W., and H.L.\ are funded by the Deutsche Forschungsgemeinschaft (DFG, German Research Foundation) -- VO 899/19-1; WI 4170/3-1; LO 418/20-1.
\end{acknowledgments}

\appendix
\section{\label{sec:appendix}Linear stability analysis}
In this appendix, we carry out a linear stability analysis to get more insights into the properties of the PFC model given by Eqs.\ \eqref{eq:apfc_SI} and \eqref{eq:apfc_SII}. For this stability analysis, we consider a homogeneous stationary state with local density $\bar{\psi}$ and vanishing local polarization. When this state is slightly perturbed, $\psi(\rb,t)$ and $\pb(\rb,t)$ can be written as 
{\allowdisplaybreaks\begin{align}%
\psi(\rb,t) &= \bar{\psi} + \delta\psi(\rb,t) \,, \label{eq:LSAI} \\
\pb(\rb,t) &= \delta\pb(\rb,t) \,, \label{eq:LSAII}      
\end{align}}%
where $\delta\psi(\rb,t)$ and $\delta\pb(\rb,t)$ are the small perturbations of the density and polarization fields, respectively. 
Inserting Eqs.\ \eqref{eq:LSAI} and \eqref{eq:LSAII} into Eqs.\ \eqref{eq:apfc_SI} and \eqref{eq:apfc_SII} and subsequent linearization with respect to the perturbations results in the equations
{\allowdisplaybreaks\begin{align}%
\begin{split}%
\partial_{t} \delta\psi &= \LaplaceS \big( \big(3\bar{\psi}^{2}+\TempR+(1+\LaplaceS )^{2} \big)\delta\psi \big) \\ 
&\quad\:\!\:\!-v_{0}\,\DivS\delta\pb \,, \end{split}\label{eq:LSADI}\\
\begin{split}%
\partial_{t} \delta\pb &= -C_{1}(\LaplaceDR + D_{r})\delta\pb - v_{0}\,\GradS\delta\psi 
\end{split}\label{eq:LSADII}%
\end{align}}%
that describe the initial time evolution of the perturbations. 
Next, we expand the perturbations as 
{\allowdisplaybreaks\begin{align}%
\delta\psi(\rb,t) &= \!\!\! \sum_{(l,m)\in\,\mathcal{I}_{\infty}} \!\!\! \hat{\delta\psi}_{lm}(t) Y_l^m(\rb) \,, \\
\delta\pb(\rb,t) &= \sum_{i=1}^2\sum_{(l,m)\in\,\mathcal{I}_{\infty}} \!\!\! \dpbc_{lm}^{(i)}(t) \mathbf{y}_{lm}^{(i)}(\rb) \,.
\end{align}}%
This leads to ordinary differential equations for the time-evolution of the expansion coefficients $\hat{\delta\psi}_{lm}(t)$ and $\dpbc_{lm}^{(i)}(t)$.
When defining the perturbation mode vector $\delta\boldsymbol{\hat{\Xi}}=(\hat{\delta\psi}_{lm},\dpbc_{lm}^{(1)},\dpbc_{lm}^{(2)})^{\mathrm{T}}$, these time-evolution equations can be written as
\begin{equation}
\partial_{t}\delta\boldsymbol{\hat{\Xi}} = -\boldsymbol{\mathrm{M}}\,\delta\boldsymbol{\hat{\Xi}}
\end{equation}
with the matrix $\boldsymbol{\mathrm{M}}=(M_{ij})_{i,j=1,2,3}$, whose elements $M_{ij}$ are given by 
{\allowdisplaybreaks\begin{align}%
M_{11} &= \frac{l(l+1)}{R^2}\Big(3\bar{\psi}^{2}+\TempR+\Big(1-\frac{l(l+1)}{R^2}\Big)^2\Big) \,, \\
M_{12} &= -v_{0}\frac{l(l+1)}{R} \,, \\
M_{21} &= \frac{v_{0}}{R} \,, \\
M_{22} &= M_{33} = C_{1}\Big(\frac{l(l+1)}{R^2} + D_r\Big) \,, \\
M_{13} &= M_{31} = M_{23} = M_{32} = 0 \,.
\end{align}}%
The three eigenvalues of this matrix are given by $\Lambda_{1} = M_{22}$ and $\Lambda_{2,3}=(M_{11}+M_{22} \pm\sqrt{D})/2$
with the real-valued $D=(M_{11}-M_{22})^{2}+4M_{12}M_{21}$.

We know that the homogeneous state of the model is stable when the real parts of all eigenvalues are positive and that it is unstable when at least one eigenvalue has a negative real part. Otherwise the linear stability analysis does not permit an assessment of the stability of the homogeneous state.
Taking into account that the signs of $M_{22}$ and $C_{1}$ are equal, since always $R>0$, $D_{r}>0$, and $l\geq 0$, we find the following stability criteria: The homogeneous state is 
{\begin{itemize}\itemindent-3mm
\item stable if $C_{1}>0 \,\wedge\, \forall l \,:\, M_{11}+M_{22}-\Re(\sqrt{D})>0$ and
\item unstable if $C_{1}<0 \,\vee\, \exists l \,:\, M_{11}+M_{22}-\Re(\sqrt{D})<0$.
\end{itemize}}%
\noindent Here, $\Re(\sqrt{D})$ denotes the real part of $\sqrt{D}$. In the case of an unstable homogeneous state, small perturbations grow with time and $\psi(\rb,t)$ and $\pb(\rb,t)$ become strongly inhomogeneous as in the crystalline states described in Sec.\ \ref{sec:results}.     
Regarding the unstable case, we can distinguish two situations: When $D\geq 0$, the amplitudes of the inhomogeneities grow with time, but their positions are static; in contrast, for $D<0$, traveling inhomogeneities emerge, since the eigenvalues $\Lambda_{2}$ and $\Lambda_{3}$ are then complex conjugates of each other (see Ref.\ \cite{WittkowskiSC2017} for details). This finding is highly interesting, since it is analogous to the observation of static and traveling crystals in the simulations. 

\begin{figure}[ht]
\includegraphics[width=\linewidth]{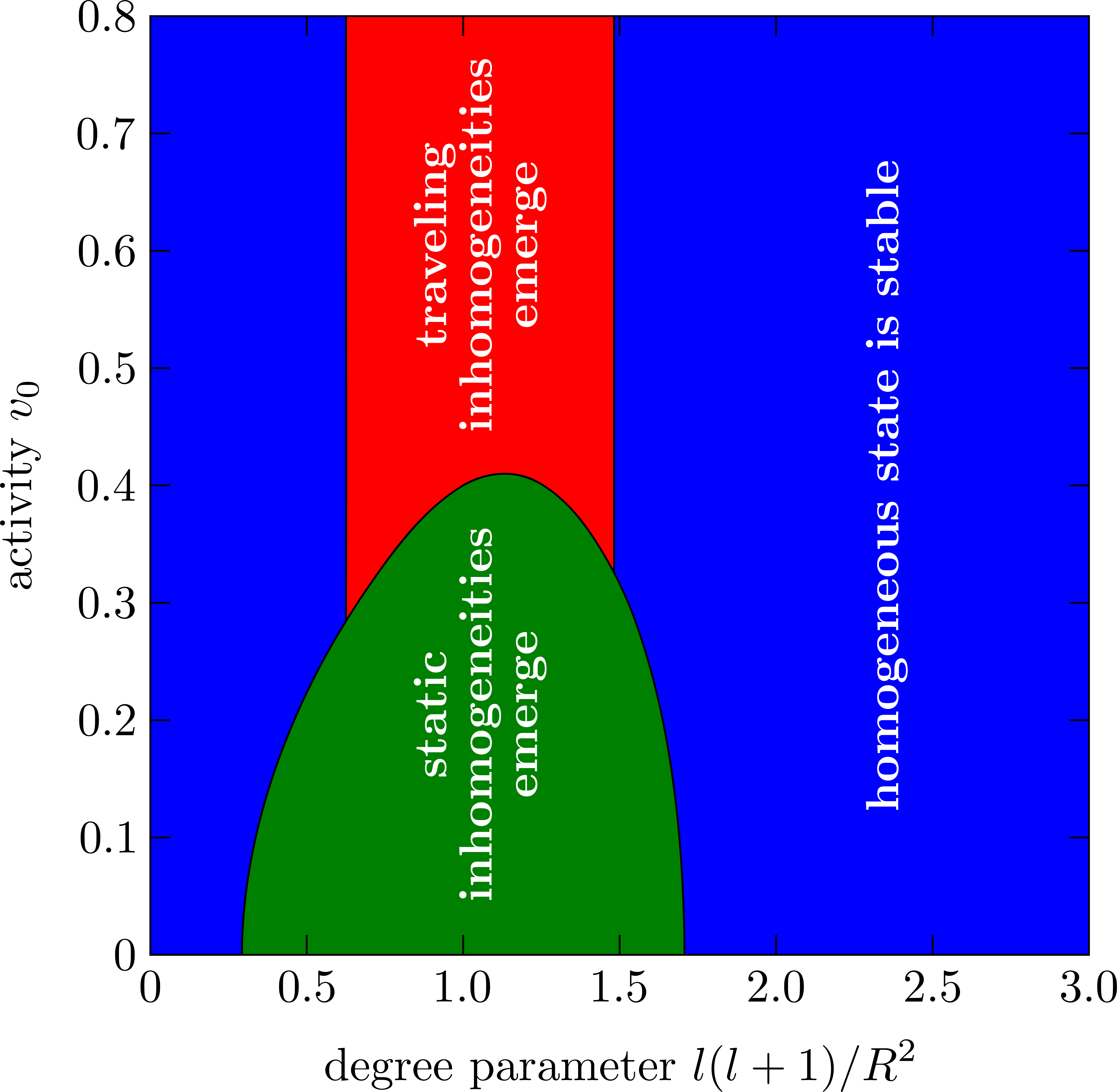}%
\caption{\label{fig12}(Color online) Stability diagram showing for various sphere radii $R$ and activities $v_{0}$ the degrees $l$ of (vector) spherical harmonic perturbations to which a homogeneous state of the PFC model given by Eqs.\ \eqref{eq:apfc_SI} and \eqref{eq:apfc_SII} with local density $\bar{\psi}$ and vanishing local polarization is stable (blue) or unstable (green or red). The emergence of inhomogeneities leads to the formation of pronounced patterns, where static inhomogeneities (green) correspond to the static crystal and traveling inhomogeneities (red) to the dynamic patterns shown in Fig.\ \ref{fig5}.}
\end{figure}

To consider this finding in more detail, we evaluated the aforementioned stability criteria for parameter values that correspond to our simulations. Remarkably, in these stability criteria the (vector) spherical harmonics degree $l$ and the sphere radius $R$ occur always together as the degree parameter $l(l+1)/R^{2}$. Therefore, we varied $l(l+1)/R^{2}$ and $v_{0}$ and chose the other parameters as in Tab.\ \ref{tab1}. This yields the stability diagram presented in Fig.\ \ref{fig12}.
It shows that for all considered activities $v_{0}$, the homogeneous state of the PFC model given by Eqs.\ \eqref{eq:apfc_SI} and \eqref{eq:apfc_SII} is unstable to (vector) spherical harmonic perturbations whose degree $l$ is within a certain range of values. This is in accordance with the fact that we observed the formation of an inhomogeneous state for all parameter combinations considered in this work. The band of degrees $l$ associated with unstable modes depends on $R$ in such a way that $l(l+1)/R^{2}$ is constant for corresponding modes in systems with different $R$. Hence, the values of $l$ associated with unstable modes increase with $R$. This is reasonable, since the emerging inhomogeneities are subject to the fixed lattice constant of $2\pi$ preferred by the model. 
Furthermore, the stability diagram shows that the emerging inhomogeneities are static for small $v_{0}$ and traveling for large $v_{0}$. This is in line with the observation of the activity threshold $v_{\mathrm{th}}$ in Fig.\ \ref{fig6}.  
In the stability diagram in Fig.\ \ref{fig12}, the activity threshold is the smallest value of $v_{0}$ for which a positive integer $l$ associated with an unstable mode exists. 
Therefore, by simultaneously solving the equations $M_{11}+M_{22}=0$ and $D=0$ with respect to $l(l+1)/R^{2}$ and $v_{0}$ and by choosing the solution with the smallest positive $v_{0}$, we calculated the coordinates $(l(l+1)/R^{2},v_{0})=(\alphathmin,\vthmin)\approx (0.63,0.28)$ of the point in the left bottom corner of the red area in the stability diagram. The activity value $\vthmin\approx 0.28$ is the threshold activity $v_{\mathrm{th}}$ for all $R$ for which the equation $l(l+1)/R^{2}=\alphathmin$ has a positive integer solution for $l$. For all other not too small $R$, the value of $v_{\mathrm{th}}$ is slightly larger, since it corresponds to the lowest point on the border between the green and red areas in Fig.\ \ref{fig12} for which $l$ is integer. An exception constitute only too small radii $R\lesssim 2$, for which the equation has no positive integer solution. This means that for $R\gtrsim 2$, the activity threshold $v_{\mathrm{th}}$ has only a weak dependence on $R$. Its values vary between $\vthmin$ and slightly larger values, where the deviations from $\vthmin$ decrease for growing $R$ and asymptotically vanish for $R\to\infty$. For the radii $R=20$ and $R=80$ used in our simulations, the activity threshold is $v_{\mathrm{th}}\approx 0.31$ and $v_{\mathrm{th}}\approx 0.29$, respectively. This is in very good agreement with the threshold value $v_{\mathrm{th}}\approx 0.3$ in Fig.\ \ref{fig6} and its apparent independence of $R$.

\bibliographystyle{apsrev4-1}
\bibliography{refs}

\end{document}